\documentclass[doublecol,linenumbers]{epl2} 

\newcommand{\bra}[1]{\langle #1 \vert}
\newcommand{\ket}[1]{\vert #1 \rangle}

\usepackage{amsmath}

\title{Entanglement generation between distant parties 
via disordered spin chains}
\shorttitle{Title} 

\author{G. M. A. Almeida\inst{1} \and F. A. B. F. de Moura\inst{1} \and M. L. Lyra\inst{1}}

\institute{                    
  \inst{1} Instituto de F\'{i}sica, Universidade Federal de Alagoas, 57072-900 Macei\'{o}, AL, Brazil
}
\pacs{03.65.Ud}{Entanglement and quantum nonlocality (e.g. EPR paradox, Bell's inequalities, GHZ states, etc.)}
\pacs{72.15.Rn}{Localization effects (Anderson or weak localization)}
\pacs{03.67.Hk}{Quantum communication}

\abstract{
We study the emergence of bipartite entanglement 
between a pair of spins weakly connected to the ends of a 
linear disordered $XY$ spin-1/2 channel.
We analyze how their concurrence responds to 
structural and on-site fluctuations 
embodied by long-range spatially-correlated sequences.
We show that the end-to-end entanglement 
is very robust against 
disorder and asymmetries in the channel
provided that the degree of correlations are strong enough and both 
entangling parties 
are tuned accordingly. 
Our results offer further alternatives in the design of
stable quantum communication protocols via imperfect channels.}

\begin{document}

\maketitle

\section{\label{sec1}Introduction}

Transmitting quantum states and establishing entanglement between
different quantum processing units are essential ingredients
towards the implementation of large-scale quantum computing \cite{nielsenbook}.
In this context, a promising approach relies on using solid-state devices such as spin chains 
with pre-engineered interactions as quantum channels \cite{bose03} for quantum
communication protocols.
In such, the information is usually encoded locally 
and after proper channel initialization and subsequent time evolution, 
the initial state (say, a qubit) can be retrieved at the desired location, or
entanglement can be created during the process.
%
The key point is to devise the protocols whose working principle
is based solely upon \textit{how} chain is engineered, avoiding the need
of external control as much as possible. 
%
After the overall concept was put forward by Bose in Ref. \cite{bose03},
much effort has been devoted 
to find out further schemes for performing quantum-state transfer (QST) 
\cite{christandl04,plenio04,wojcik05,wojcik07,li05,huo08, apollaro12, longhi14, lorenzo15, longhi16, almeida16}
and entanglement creation/distribution
protocols \cite{amico04,plastina07,venuti06, gualdi11, estarellas17, almeida17-1}. 

The versatility offered by spin chains comes with a price though. 
The lack of dynamical control 
implies that the manufacturing process of the chain must 
be very
accurate. Otherwise, the appearance of imperfections (e.g. disorder) should
compromise the desired output. 
There lies the importance of evaluating the robustness 
of such quantum communication protocols
in the presence of noise 
\cite{dechiara05,burgarth05,tsomokos07, petrosyan10,yao11,zwick11,ashhab15, almeida17-1}. 

In this letter,
we go along that direction and address the influence of disorder in 
weakly-coupled spin models \cite{wojcik05, wojcik07, li05}. 
In the QST framework, 
a pair of spins are perturbatively connected to, say, 
each end of a $XY$ spin-$1/2$ chain thereby
spanning a decoupled Hamiltonian involving both spins only or 
added with a given normal mode of the channel once their frequency goes in resonance
with it \cite{wojcik07}.
%
%
Ideally, that is, in a noiseless channel, this decoupling process yields the appearance of maximally entangled
Bell-type eigenstates 
which are responsible for Rabi-like
oscillations 
between the sender and receiver spins
thus allowing for high-fidelity QST performances \cite{almeida16}.
%
%
Random fluctuations, on the other hand, will act on the channel by
shuffling the spectrum, promoting localization, and thus destroying the mirror symmetry 
of every eigenstate of the system \cite{yao11}.
%
When correlated fluctuations -- say, spatially dependent -- are present
the scenario is rather different. For instance, it was shown that short-range
correlations in the disorder distribution promotes the breakdown of Anderson localization
in 1D models \cite{phillips91}. The effect of long-range correlations
is even more dramatic as it induces a metal-insulator transition with sharp
mobility edges \cite{demoura98, adame03}. The coexistence between localized and delocalized states hence provides a rich set of dynamical 
regimes to explore \cite{adame03,lima02,demoura02}.
For example, it has been reported \cite{almeida17-1} that the emergence of correlated fluctuations
in the local magnetic fields
is capable of enhance the distribution of entanglement 
in $XY$ spin chains.

Here we investigate how long-range correlated diagonal (local magnetic fields) and off-diagonal (spin couplings) disorder affect the stability
of the entanglement developed between 
two distant spins, both being perturbatively coupled to the disordered channel.
We consider fluctuations which follow a power-law spectrum of the form 
$S(k)\propto 1/k^{\alpha}$, with $k$ being the corresponding wave number
and $\alpha$ is an exponent that characterizes the \textit{degree} 
of those correlations with $\alpha = 0$ standing for the
uncorrelated case. 
For high-enough values of $\alpha$ we find that the end-to-end concurrence reaches about
its maximum value despite the fact that significant amounts
of disorder are still present. We further discuss
what kind of profile a given channel must
have in order to mediate entanglement in the presence of imperfections.
The emergence of a set of delocalized states in the 
middle of the spectrum when $\alpha > 2$
\cite{demoura98}
becomes crucial in generating an
effective resonant coupling between both distant parties.
Our findings settle a resilient framework
for carrying out long-distance quantum communication protocols
through highly disordered quantum channels.
%
   

\section{\label{sec2}Model Hamiltonian}

Thorough the paper we consider a one-dimensional isotropic XY spin-$1/2$ chain
with open boundaries featuring $N+2$ spins labelled by $i=0,1,2,\ldots,N,N+1$, 
with spins $0$ and $N+1$
denoting the parties aimed to get entangled. 
The full Hamiltonian of the system reads ($\hbar = 1$)
\begin{equation} \label{Hfull}
\hat{H} = \sum_{i=0}^{N+1}\dfrac{\omega_{i}}{2}(\hat{1}-\hat{\sigma}_{i}^{z})-\sum_{i=0}^{N}\dfrac{J_{i}}{2}(\hat{\sigma}_{i}^{x}\hat{\sigma}_{i+1}^{x}
+\hat{\sigma}_{i}^{y}\hat{\sigma}_{i+1}^{y}), 
\end{equation}
where $\hat{\sigma}_{i}^{x,y,z}$ comprise the Pauli operators for the $i$-th spin, 
$\omega_{i}$ is the local magnetic field, 
and $J_{i}$ is the nearest-neighbor
exchange coupling rate (all those being real parameters). 
For simplicity, we set the above Hamiltonian to be expressed 
in terms of an arbitrary energy unit $J=1$.
Hereafter we relabel $0\rightarrow s$ and $N+1 \rightarrow r$  
and assume those to be 
weakly coupled to each end of the channel, hence fixing $J_{0}\rightarrow g_{s}$ and
$J_{N}\rightarrow g_{r}$, both being much smaller than any other $J_{i}$,
and free of imperfections alongside $\omega_{s}$ and $\omega_{r}$.
Therefore, we consider disorder to take place in the channel only (that is, from spins $1$ to $N$). 
This is a reasonable assumption in the sense that spins $s$ and $r$ are 
the only components of the system supposed 
to feature a higher degree of control due to the need of performing 
state preparation and read-out protocols on them. We will specify the disorder
distribution later on. 

Note that $\left[ \hat{H}, \sum_{i} \hat{\sigma}_{i}^{z}\right] =0$ and hence
Hamiltonian (\ref{Hfull}) is made up by 
independent blocks with fixed number of excitations. 
In standard QST protocols \cite{bose03}, one wishes
to transmit the state of a single qubit through
an initially polarized channel such that 
$\ket{\Psi (0)} = \ket{\phi}_{s}\ket{\downarrow_{1},\ldots , \downarrow_{N}, \downarrow_{r}}$ with 
$\ket{\phi}_{s} = a \ket{\downarrow_{s}}+ b \ket{\uparrow_{s}}$. The goal
is to achieve 
$\ket{\Psi (t)} = \ket{\downarrow_{s},\downarrow_{1}, \ldots , \downarrow_{N}}\ket{\phi}_{r}$
in a given time $t$. 
The whole process takes place in subspaces with none and single
flipped spins, the latter being the only component
which actually evolves in time. 
We thus carry out 
our investigation on the subspace spanned
by $\ket{i}\equiv \ket{\downarrow_{s},\downarrow_{1},\downarrow_{2}, \ldots, \downarrow_{i-1},\uparrow_{i},\downarrow_{i+1}, \ldots}$.

In this work we consider the presence of 
\textit{static} disorder in the channel,
taking place either 
in the local magnetic fields or in the spin couplings. 
We model that by sequences featuring long-range correlations  
following a power-law spectrum of the form $S(k)\propto 1/k^{\alpha}$ as generated from
\cite{demoura98, adame03}
\begin{equation} \label{disorder}
\omega_{n}, J_{n} = \sum_{k=1}^{N/2}k^{-\alpha/2}
\mathrm{cos}\left( \dfrac{2\pi n k}{N} + \phi_{k} \right),
\end{equation} 
where $n=1,\ldots,N$, $\lbrace \phi_{k} \rbrace$ are random
phases uniformly distributed in the interval $\left[0,2\pi \right]$ 
and $\alpha$ 
stands for the degree of those underlying correlations. 
We remark that 
the above distribution possess no typical length scale, 
as it does many natural stochastic series \cite{bak96}.
Uncorrelated disorder, i.e. white noise, is recovered when $\alpha=0$. 
Indeed, $\alpha$ is directly related to the so-called Hurst exponent $H$ \cite{fractalsbook} through $H = (\alpha -1)/2$ which characterizes
self-similariry of a given series. The sequence 
spanned by Eq. (\ref{disorder}) becomes
nonstationary when $\alpha > 1$ and is said to be 
persistent (anti-persistent) when $\alpha > 2$ ($\alpha < 2$).
The $\alpha=2$ case is where the sequence corresponds exactly to the trace of 
the Brownian motion. 
In our calculations, we set the disorder sequence to have 
zero mean and unit variance, $X_{n} \rightarrow 
\left( X_{n}-\langle X_{n}\rangle \right) / \sqrt{\langle X_{n}^2\rangle-\langle X_{n}\rangle^2}$, with $X_{n}$ representing any stochastic variable. 
When considering disorder 
in the spin couplings, we also recast $J_{n}\rightarrow J_{n}+4.5$ in order
to guarantee that none of them vanishes.

\section{\label{sec3}Perturbation theory}

Now, we proceed by carrying out a perturbative approach in order to 
find an equivalent Hamiltonian that effectively decouples
the sender/receiver pair from the rest of the channel.
To do so, we go along the procedure used in Ref. \cite{wojcik07}.
The first step is to gather all the channel
local states, $\ket{1}$ to $\ket{N}$, and combine them so as to
diagonalize Hamiltonian (\ref{Hfull}) 
with 
$g_{s}=g_{r} = 0$, thus obtaining the channel normal modes $\lbrace \ket{E_{k}} \rbrace$ and
their associated frequencies $\lbrace E_{k} \rbrace$, which we assume to be nondegenerate (we do not need to worry about their exact form at this point).
Wiring up 
spins $s$ and $r$ to the channel we express the full Hamiltonian
in a very convenient form 
$\hat{H} = \hat{H}_{0}+\hat{V}$, where
\begin{align}
\hat{H}_{0} = \omega_{s}\ket{s}\bra{s} +  \omega_{r}\ket{r}\bra{r}  + \sum_{k}E_{k} \ket{E_{k}}\bra{E_{k}}, \\ 
\hat{V} = \epsilon \sum_{k} \left( g_{s}a_{sk} \ket{s}\bra{E_{k}} + g_{r}a_{rk} \ket{r}\bra{E_{k}} + \mathrm{H.c.} \right),
\end{align} 
with $a_{sk}\equiv\langle 1 \vert E_{k} \rangle$, and 
$a_{rk}\equiv\langle N \vert E_{k} \rangle$ being real-valued coefficients.
Note that we have introduced a perturbation parameter $\epsilon$
in order to assure that spins $r$ and $s$ do not disturb the band.
Therein, we highlight two possibilities, namely (i) both spins
are out of resonance with all the normal modes of the channel, or,
(ii) they eventually meet one of those such that $\omega_{s}=\omega_{r}=E_{k'}$.
Let us treat each case separately. 

In the first scenario (i), we define $\hat{H}_{\mathrm{eff}} = e^{i\hat{S}}\hat{H}e^{-i\hat{S}}$ with $\hat{S}$ being a Hermitian operator with entries
$\langle \nu \vert \hat{S} \vert E_{k} \rangle = i\epsilon g_{\nu}a_{\nu k} /(E_{k}-\omega_{\nu})$, where $\nu \in \lbrace s,r \rbrace$ (see \cite{wojcik07, li05}).
Expanding $\hat{H}_{\mathrm{eff}}$ up to second order in $\epsilon$ we get
\begin{equation} \label{Heff}
\hat{H}_{\mathrm{eff}} = \hat{H}_{0}+\hat{V}+ i[\hat{S},\hat{H}_{0}]+ i[\hat{S},\hat{V}] +\dfrac{i^{2}}{2!}[\hat{S},[\hat{S},\hat{H_{0}}]] + O(\epsilon^{3}). 
\end{equation}
Due to the choice of $\hat{S}$, the first order terms in the above expression
vanish, $\hat{V}+i[\hat{S},\hat{H}_{0}] = 0$. Then,
by inspecting Eq. (\ref{Heff}),  
we have effectively decoupled the outer spins from the channel, that is
$\hat{H}_{\mathrm{eff}} = \hat{H}_{sr} \oplus \hat{H}_{\mathrm{ch}}$, 
where \cite{wojcik07}
\begin{equation}\label{Heff2}
\hat{H}_{sr}= h_{s}\ket{s}\bra{s}+h_{r}\ket{r}\bra{r} -J'\left(\ket{s}\bra{r}+ \mathrm{H.c.}\right),
\end{equation} 
thus
describing a two-level system with effective local potentials and coupling given by, respectively, 
\begin{equation}\label{weff}
h_{\nu} = \omega_{\nu}-\epsilon^{2} g_{\nu}^{2} \sum_{k}\dfrac{|a_{\nu k}|^{2}}{E_{k}-\omega_{\nu}},
\end{equation}
\begin{equation}\label{Jeff}
J' = \dfrac{\epsilon^{2}g_{s}g_{r}}{2}\sum_{k}\left( \dfrac{a_{sk}a_{rk}}{E_{k}-\omega_{s}} + \dfrac{a_{sk}a_{rk}}{E_{k}-\omega_{r}} \right).
\end{equation}
The eigenstates of Hamiltonian (\ref{Heff2}) can be easily handled out analytically, yielding
\begin{equation} \label{state2}
\ket{\psi^{\pm}} = \dfrac{2J'\ket{s}+(\Delta \pm \Omega)\ket{r}}{\sqrt{(\Delta \pm \Omega)^{2}+4J'^{2}}},
\end{equation}
where $\Delta \equiv h_{s}-h_{r}$ is the effective detuning and
$\Omega = \sqrt{\Delta^{2}+4J'^{2}}$ is their corresponding Rabi-like frequency.
In order to achieve a maximally entangled state, one then needs $\Delta = 0$ (or, at least, $\Delta \gg J'$) which
results in $\ket{\psi^{\pm}} = (\ket{s}\pm\ket{r})/\sqrt{2}$.

In case (ii) where both outer spins achieves resonance
with a particular mode of the channel (say, labelled by $k'$), leaving the rest of the band untouched so that we can safely neglect off-resonant, fast-rotating 
interactions, an effective three-level Hamiltonian can be obtained \cite{wojcik07},
\begin{equation} \label{Heff3}
\hat{H}_{sk'r} = \epsilon \left( g_{s}a_{sk'} \ket{s}\bra{E_{k'}} + g_{r}a_{rk'} \ket{r}\bra{E_{k'}} + \mathrm{H.c.} \right),
\end{equation}
where we have shifted the local frequencies (diagonal terms) $E_{k'} \rightarrow 0$ 
with no loss of generalization. Again, diagonalizing the above Hamiltonian
is quite straightforward. Particularly, we are interested in the zero eigenstate
$\ket{\psi_{0}}\equiv (d_{1},d_{2},d_{3})$ which fulfils  
$\hat{H}_{sk'r}\ket{\psi_{0}} = 0$. Thereby, one must have $d_{2} = 0$ (no component in $\ket{E_{k'}}$) and
$d_{1}g_{s}a_{sk'}+ d_{3}g_{r}a_{rk'} = 0$. After proper normalization, the eigenstate reads
\begin{equation} \label{state3}
\ket{\psi_{0}} = \dfrac{\ket{s}-\eta\ket{r}}{\sqrt{1+\eta^{2}}},
\end{equation}
where $\eta \equiv g_{s}a_{sk'}/g_{r}a_{rk'}$ accounts for the balance between 
both effective couplings. 
Obviously, when $\eta = \pm 1$ the zero mode becomes fully entangled. 

Note that up to now we have not made any specific 
assumptions towards the profile of the channel eigenstates so that the above formulas 
hold for any arbitrary network and provide us great insight 
about how the entanglement between the
outer spins $s$ and $r$ depend upon the spectrum properties of the channel. 

\section{\label{sec4}Long-distance entanglement}

A powerful tool to quantify bipartite entanglement
between two qubits in an arbitrary mixed state is the so-called concurrence \cite{wootters98}. Consider an arbitrary quantum state written 
on the computational basis,
$\ket{\psi} = \sum_{i}d_{i}\ket{i}$, with $d_{i}$ being, in general, 
a complex coefficient. The input information we need to proper characterize entanglement 
between a given pair of spins $i$ and $j$ is found in their corresponding
reduced density matrix $\rho_{ij}$ defined in basis $\lbrace \ket{\downarrow_{i}\downarrow_{j}},\ket{\uparrow_{i}\downarrow_{j}},\ket{\downarrow_{i}\uparrow_{j}},\ket{\uparrow_{i}\uparrow_{j}} \rbrace$, 
which is obtained by tracing out
all the remaining sites. Concurrence is then defined as  
$C(\rho_{ij}) = \mathrm{max}\lbrace 0, \sqrt{\lambda_{1}}-\sqrt{\lambda_{2}}-\sqrt{\lambda_{3}}-\sqrt{\lambda_{4}} \rbrace $ \cite{wootters98}, 
where $\lbrace \lambda_{i} \rbrace$ are the eigenvalues, in decreasing order, of the non-Hermitian matrix $\rho_{ij} \tilde{\rho}_{ij}$, with
$\tilde{\rho}_{ij} = (\hat{\sigma}_{y}\otimes \hat{\sigma}_{y})\rho_{ij}^{*}(\hat{\sigma}_{y}\otimes \hat{\sigma}_{y})$. 
In the single-excitation manifold, straightforward calculations lead to (for details, see \cite{amico04})
\begin{equation}\label{c_ij}
C_{i,j} \equiv C(\rho_{ij}) = 2\vert d_{i}d_{j}^{*}\vert,
\end{equation}
which gives $C_{i,j} = 0$ for non-entangled (separable) spins and
$C_{i,j} = 1$ for maximally-entangled parties. 

By inspecting Eqs. (\ref{state2}) and (\ref{state3}) 
from the previous section, one gets
$C_{s,r} = 2|J'\Omega^{-1}|= 2[(\Delta/J')^{2}+4]^{-1/2}$ for
the effective two-level description given by
$H_{sr}$ [Eq. (\ref{Heff2})]
and $C_{s,r} = 2|\eta|(1+\eta^{2})^{-1}$ for the three-level regime expressed by
$H_{sk'r}$ [Eq. (\ref{Heff3})].
Therefore, we verify that the quality of 
entanglement between spins $s$ and $r$ will be ultimately dictated 
by the quantities $\Delta/J'$ and $\eta$, 
depending on the interaction regime we are dealing with.
Setting $g_{s}=g_{r}$
and $\omega_{s} = \omega_{r}$, we note that for noiseless mirror-symmetric channels the conditions 
for maximal entanglement are readily fulfilled, since  
$|a_{sk}| = |a_{rk}|$ $\forall$ $k$, which gives $\Delta = 0$, $|\eta| = 1$, and thus $C_{s,r} = 1$. 

Fluctuations in the parameters of the chain can then seriously damage
the above (very suitable) scenario. 
Uncorrelated disorder ($\alpha = 0$) in 1D hopping models is known for
induce the phenomenon Anderson localization \cite{anderson58} when 
every eigenstate of the system acquires the form
$\langle x\vert E_{k}\rangle \sim e^{-\frac{|x-x_{0}|}{\xi_{k}}}$, thus 
becoming exponentially localized around a given position $x_{0}$ with
$\xi_{k}$ accounting for length of localization.
As a consequence, it becomes extremely unlikely to
find
eigenstates $\ket{E_{k}}$ featuring 
comparable amplitudes $a_{sk}$ and $a_{rk}$
anywhere in the spectrum. In principle, 
one could fight against that by locally tuning 
either $g_{\nu}$ and/or $\omega_{\nu}$ to compensate such distortion \cite{yao11}.
In the effective three-level regime described by $H_{sk'r}$, that should work well since
all we need is $\eta = g_{s}a_{sk'}/g_{r}a_{rk'} \approx  1$ given a \textit{fixed} frequency $E_{k'}$, otherwise we must also reset $\omega_{\nu}$ 
to find another mode to tune with, for each disorder configuration.  
In the effective two-level case, though, the situation is more subtle.
First of all, note that the ratio $\Delta/J$ does not depend upon
a single channel eigenstate but on the entire spectrum [see Eqs. (\ref{weff}) and (\ref{Jeff})] weighted by the inverse of the ``distance'' between $E_{k}$ and $\omega_{\nu}$. In addition, in this case there is no much liberty in manipulating
$g_{\nu}$ since it might lead to a mixing between the 
channel and sender/receiver subspaces,
thus invalidating Hamiltonian (\ref{Heff2}).
Despite all that, since there is no way to exactly predict a given
disorder configuration,  
it should be much more preferable to fix $\omega_{\nu}$,
$g_{\nu}$ and make sure that, statistically, the protocol yields
successful outcomes up to a given user-defined threshold. 
This also rules out the need of additional resources.

 
We are now about to discuss the effects of long-range correlated disorder
in the generation of entanglement.  
Let us start by considering noise on the local magnetic fields distribution $\lbrace \omega_{n} \rbrace$ [given by Eq. (\ref{disorder})] and fixing $J_{n} = 1$ (in units of $J$).
Note that regardless of the parity of the chain size, disorder shuffles every natural
frequency of the channel and so there will be no unique level
to satisfy Hamiltonian (\ref{Heff3}).
For this reason, in the case of on-site disorder we 
only address the likelihood of entanglement taking place in the effective two-level  scenario [Eq. (\ref{Heff2})].  
%

%
\begin{figure}[t!] 
\includegraphics[width=0.4\textwidth]{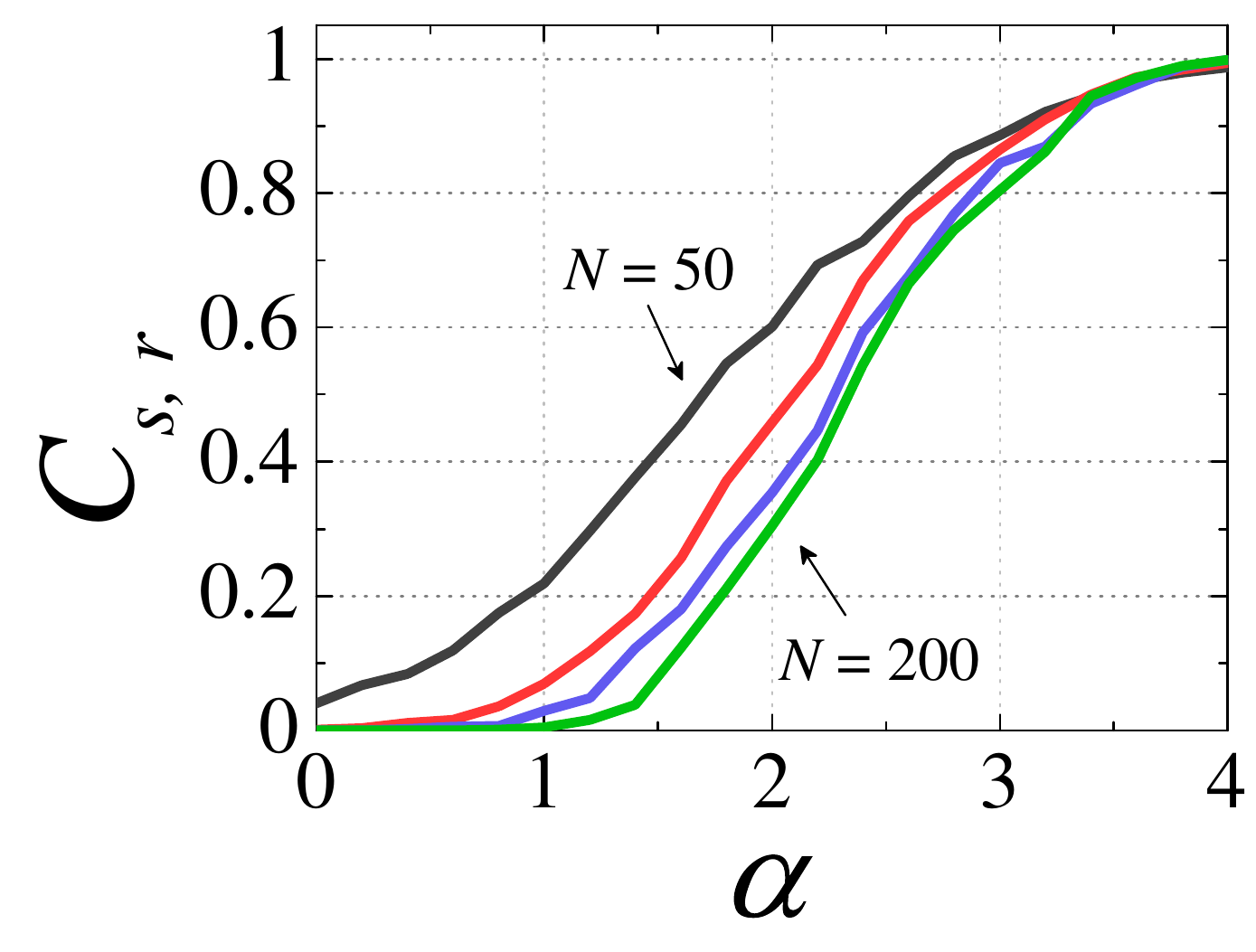}
\caption{\label{fig1} (Color online) End-to-end concurrence for the 
effective two-level case,  
$C_{s,r} =  2[(\Delta/J')^{2}+4]^{-1/2}$,
versus $\alpha$ averaged over 500 independent realizations of on-site disorder 
for various channel sizes $N=$ $50$, $100$, $150$, and $200$. 
We set $\omega_{r} =\omega_{s} = 0 $, $g_{s} = g_{r}$ and $J_{n}/J=1$ (channel spin-coupling strengths).
The input parameters of the channel, $a_{\nu k}$ and $E_{k}$,
were obtained directly from exact
numerical diagonalization of Hamiltonian (\ref{Hfull}) without
the outer spins. 
}
\end{figure}
%
%

%
\begin{figure}[t!] 
\includegraphics[width=0.45\textwidth]{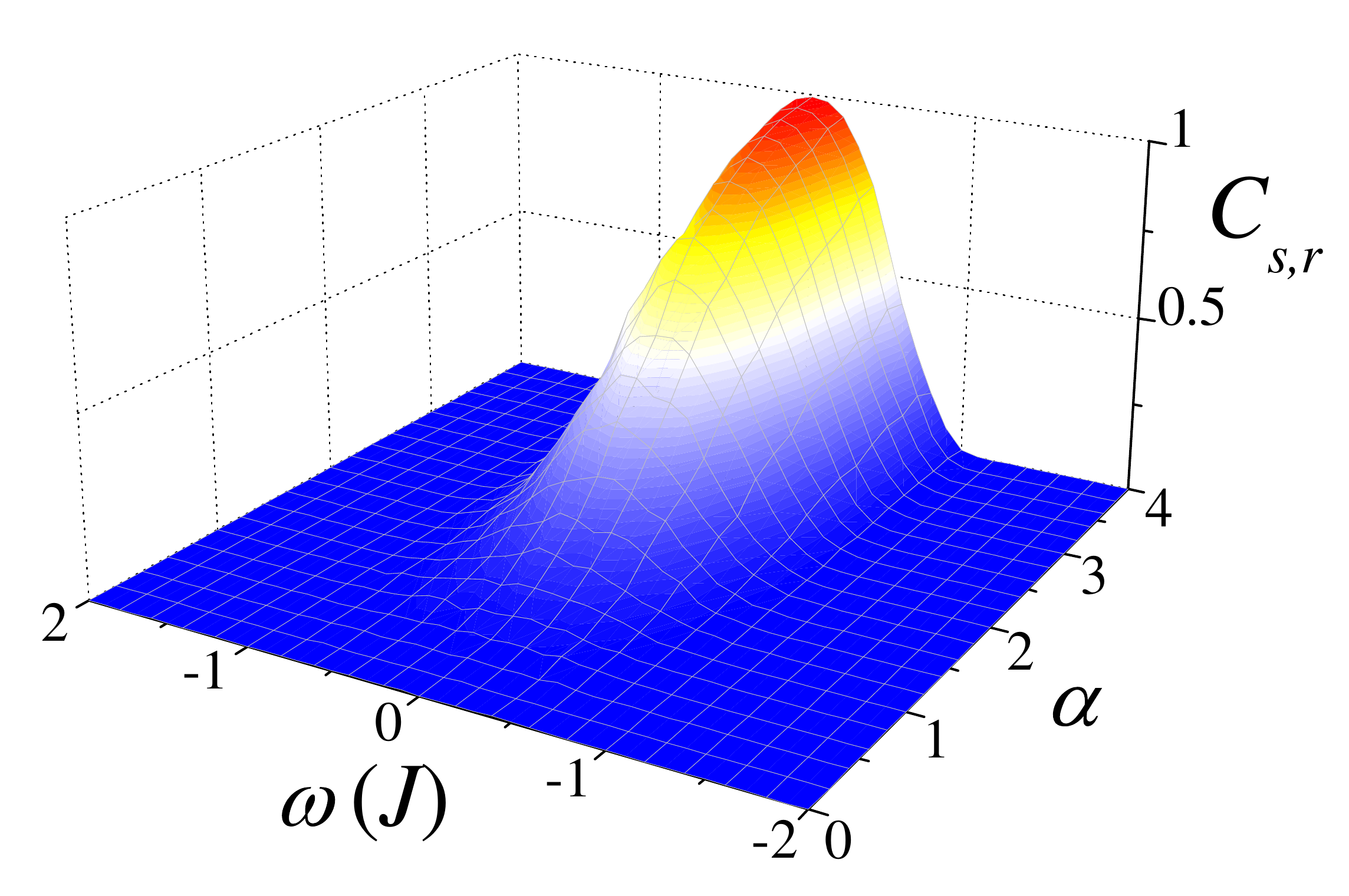}
\caption{\label{fig2} 
(Color online) 
End-to-end concurrence for the 
effective two-level case,  
$C_{s,r} =  2[(\Delta/J')^{2}+4]^{-1/2}$,
for varying $\alpha$ and frequency 
$\omega = \omega_{r} = \omega_{s}$ averaged over 500 independent 
realizations of on-site disorder 
for $N=100$, $g_{s}=g_{r}$, and $J_{n}/J=1$.
Therein, we certify that maximal entanglement can only 
be achieved when tuning spins $s$ and $r$ around the center of the band.
}
\end{figure}

In Fig \ref{fig1} we show the resulting behavior of the disorder-averaged concurrence
as a function of the degree of correlations $\alpha$ for different sizes of the channel, $N$. 
It starts off with a very poor figure of merit
for lower values of $\alpha$, as expected, and reaches about its maximum,
$C_{s,r}\approx 1$, when
$\alpha=4$. As $N$ increases -- thus weakening finite-size effects --
we also note that
the concurrence begins to suddenly build up after $\alpha=1$. 
This is related to the fact that the series generated by Eq. (\ref{disorder})
becomes nonstationary, preceding the appearance of  
a set of delocalized states around the center of the band,
induced by the persistent character of the series
when $\alpha > 2$ \cite{demoura98}. 
This region of the spectrum thus offers a suitable ground for creating 
entanglement since the corresponding channel eigenstates $\lbrace\ket{E_{k}}\rbrace$ are expected to feature more balance between $a_{sk}$ and $a_{rk}$, 
this way increasing the possibilities of having $\Delta/J' \ll 1$. 
The outer parts of the band, still composed 
by strongly-localized modes, 
have a much weaker influence on that since the terms inside the sum
in Eqs. (\ref{weff}) and (\ref{Jeff}) decreases following $E_{k}^{-1}$.
Because of that, the sender/receiver local frequencies 
must be set as close as possible 
to the center of the band (as in Fig. \ref{fig1}). 
If they do not, Fig. \ref{fig2} shows exactly what happens.
As we shift $\omega = \omega_{s}= \omega_{r}$ away from $0$, the concurrence
drops very rapidly since the effective coupling between both spins
becomes more sensitive to the imperfections and asymmetries of the channel.   

%
\begin{figure}[b!] 
\includegraphics[width=0.4\textwidth]{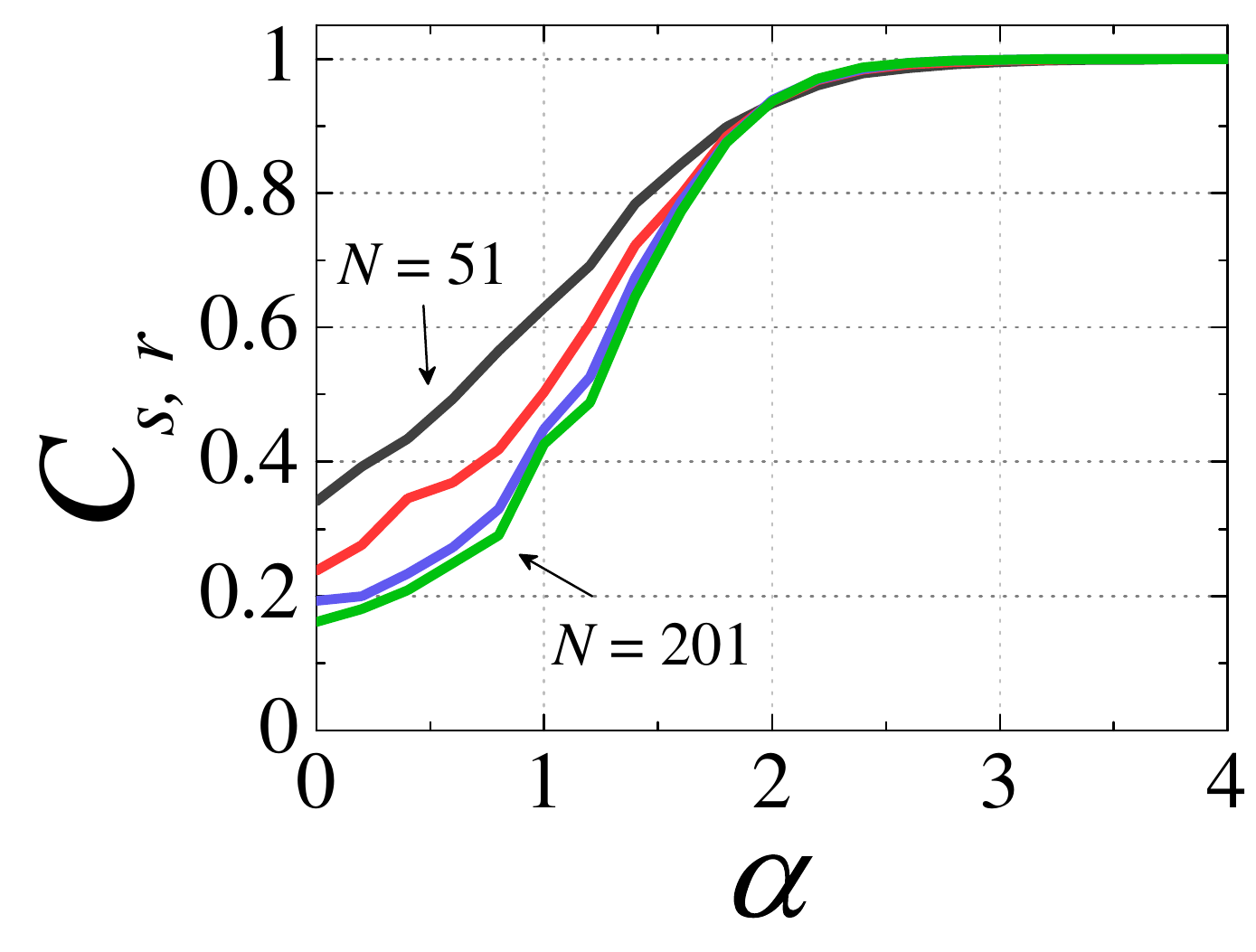}
\caption{\label{fig3} 
(Color online) End-to-end concurrence for the 
effective three-level regime,  
$C_{s,r} = 2|\eta|(1+\eta^{2})^{-1}$,
versus $\alpha$ averaged over 500 independent realizations of spin-coupling strength disorder 
for various channel sizes $N=$ $51$, $101$, $151$, and $201$. 
We set $\omega_{s} =\omega_{r} = 0 $ (matching the central anomalous mode), $g_{s} = g_{r}$, and $\omega_{n} = 0$ (channel local magnetic fields).
}
\end{figure}

Now let us move on to the case of structural disorder, i.e., 
fluctuations affecting the spin-coupling strengths of the channel, $\lbrace J_{n}\rbrace$. 
First, we must have in mind that this kind of disorder, despite breaking the spatial mirror symmetry of the system, it preserves particle-hole symmetry 
meaning that $E_{k} = -E_{-k}$ and $|a_{\nu k}| = |a_{\nu -k}| $ 
for every eigenstate, considering an even $N$.
Therefore, in the off-resonant two-level interaction regime 
we trivially get $\Delta/J' = 0$ and so $C_{s,r} = 1$
provided $\omega_{\nu} = 0$ and $g_{\nu}$ is small enough, so as to justify
the two-level approximation [Eq. (\ref{Heff2})].  

Another very relevant aspect of nearest-neighbor off-diagonal disorder
in tight-bindings models
is the emergence
of an anomalous mode at the very center of the band 
featuring a diverging localization length. 
Indeed,
the density of states shows a logarithmic singularity in this region \cite{inui94}.
For odd values of $N$ we thus 
get a \textit{fixed} energy level
$E_{k'} = 0$ for every realization of disorder
%
Thereby,  
this time we focus on the effective three-level scenario, Hamiltonian (\ref{Heff3}), with
$\omega_{s} =\omega_{r} = E_{k'} $.

Figure \ref{fig3} shows the behavior of concurrence with $\alpha$. 
We note that uncorrelated disorder ($\alpha = 0$)
already offers a reasonable amount of entanglement though
it diminishes with increasing $N$. 
This is due to the finiteness of the system added by
the unusual properties of the corresponding channel eigenstate at $E_{k'} = 0$
that acts as sort of a pseudo-delocalized state with a stretched
exponential envelope \cite{inui94}.
For $N=151$ and $201$, we again 
spot a pronounced increase in the concurrence when $\alpha>1$. 
This naturally comes out as a response to the emergence of delocalization 
around the center of the band. We shall, however, remark that
the regime $1<\alpha<2$ does not actually provide 
\textit{true} extended states \cite{cheraghchi05}, though the participation number
increases quite considerably \cite{assuncao11}. 
In this effective three-level (sender-channel-receiver resonant) scenario though, 
the outer entangling spins are being mediated \textit{solely} by  
the anomalous eigenstate at $E_{k'}=0$. We thus realize that
this state is very sensitive to $\alpha$ in such a way that
$\alpha > 2$ guarantees the necessary symmetry, $|\eta| \approx 1$,
to support maximal entanglement between both parties. 
We must stress, however, 
that this behavior could have also been seen if we had set
$\omega_{s}$ in resonance with some other nearby mode. In fact, although the 
anomalous eigenstate shows very peculiar features 
for uncorrelated structural disorder \cite{inui94, cheraghchi05}, 
the overall trend when $\alpha>0$ is -- similarly to what happens
in the on-site disorder case -- the appearance
of delocalized states in the neighborhood of $E_{k'}=0$, 
including 
the anomalous mode itself, with roughly the same
localization length \cite{lima02}. 

\begin{figure}[t!] 
\includegraphics[width=0.49\textwidth]{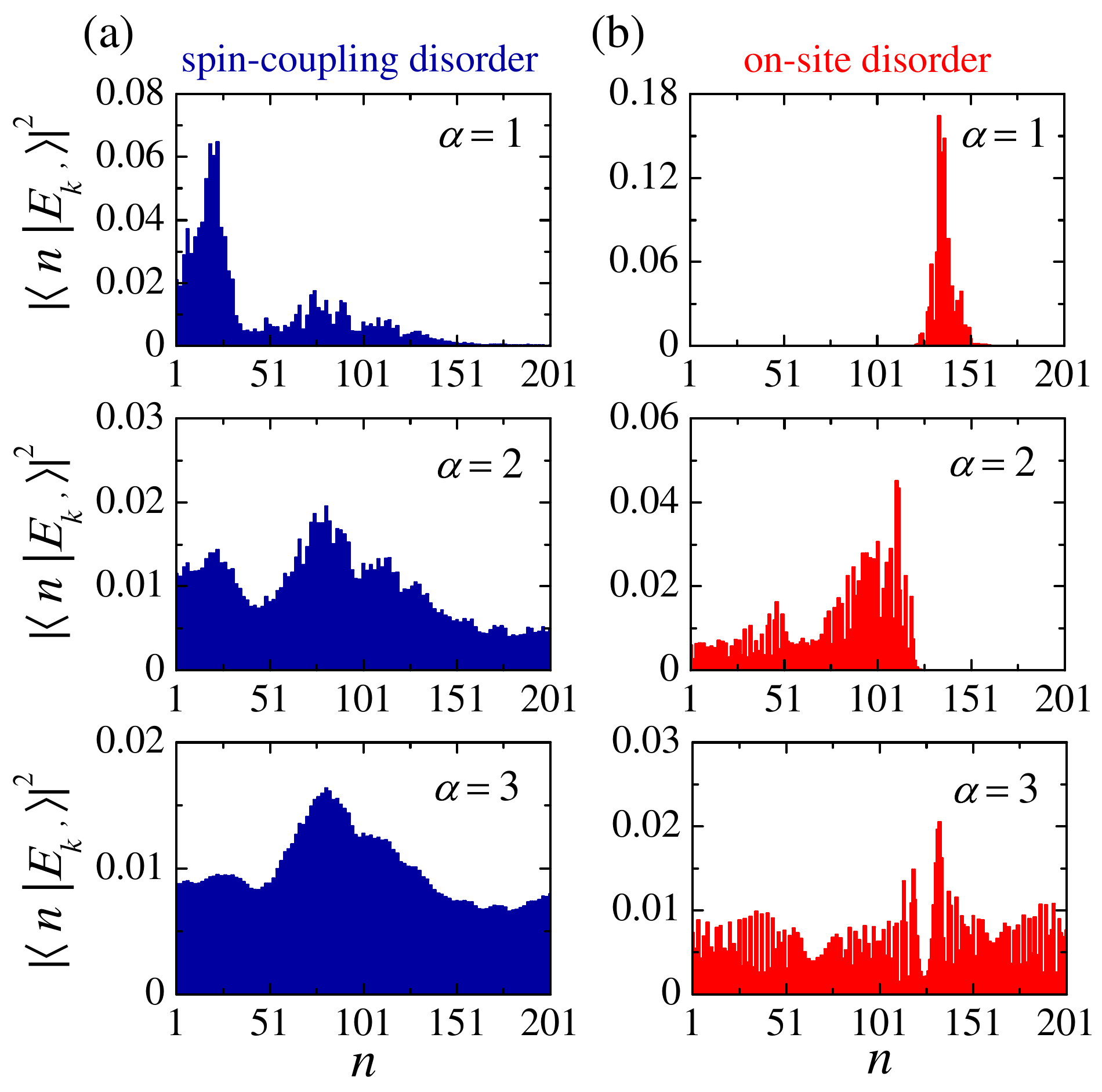}
\caption{\label{fig4} 
(Color online) Square modulus of the wavefunction $|\langle n \vert E_{k'} \rangle|^{2}$
versus $n$ for several values of $\alpha$ for a single realization of 
(a) spin-coupling disorder and (b) on-site disorder. 
The 
corresponding 
eigenstate of the channel, $\ket{E_{k'}}$, was picked from the middle of the band, $E_{k'}\approx 0$ and the size of the chain was $N=201$.
}
\end{figure}

The major difference between both diagonal and off-diagonal disorder
cases is how fast delocalization is built with $\alpha$. This can be seen indirectly 
through the stabilization of the end-to-end concurrence
by comparing Figs. \ref{fig1} and \ref{fig3}. 
In order to further illustrate that, in Fig. \ref{fig4} we show
a typical (single) realization of the wavefunction (its squared modulus)
corresponding to a given channel mode selected at $E_{k'} \approx 0$
for both kinds of disorder and $N=201$. 
As expected, in the case of spin-coupling disorder [Fig. \ref{fig4}(a)] the wavefunction features 
a much larger localization length than the on-site disorder counterpart [Fig. \ref{fig4}(b)]
already for $\alpha = 1$. When $\alpha=2$, we note in Fig. \ref{fig4}(a) 
that the state shows fairly balanced overlaps 
$|a_{sk'}| = |\langle 1 \vert E_{k'} \rangle|$
and $|a_{rk'}| = |\langle N \vert E_{k'} \rangle|$,
which is crucial for having $|\eta| \approx 1$ [cf. Eq. (\ref{state3})]
and hence $C_{s,r}\approx 1$. That explains the behavior seen in Fig. \ref{fig3}.
On the other hand, the wavefunction in Fig. \ref{fig4}(b) when $\alpha=2$
still features a reminiscent localized-like profile.  
Things get more alike only at higher values of $\alpha$ as seen in the bottom
panels of Fig. \ref{fig4} for $\alpha=3$.
At this point, note that $|a_{sk'}| \approx |a_{rk'}|$ for both 
types of disorder. 
While this is responsible for yielding 
very high concurrences (see Figs. \ref{fig1} and \ref{fig3}), as long
as we depend upon a single mode of the channel such as in the effective
three-level regime [Eq. (\ref{Heff3})] and this very state happens to be
a suitable one, there is no need to worry
about the rest of the spectrum, unlike in the two-level regime where
the effective detuning between $\ket{s}$ and $\ket{r}$ is largely influenced 
by the modes lying around $\omega_{\nu}$, 
weighted by $(E_{k}-\omega_{\nu})^{-1}$ [see Eqs. (\ref{weff}) and (\ref{Jeff})].
Still, in this case one can achieve nearly maximal entanglement provided $\alpha$ is high
enough, as seen in Figs. \ref{fig1} and \ref{fig2}. 
 
%

\section{\label{sec5}Concluding remarks}

We investigated the creation of entanglement between
the two weakly-coupled ends of a disordered linear $XY$ spin-$1/2$ chain 
in two different interaction regimes, namely when both outer spins are off-resonantly coupled to the channel and when they are are put in narrow resonance with one of its natural modes.
%
In both cases, we found that quantum channels presenting long-range correlated fluctuations is capable of mediating extremely high amounts of pairwise entanglement through long distances, thus embodying a robust
platform for carrying out quantum communication tasks
in the presence of imperfections. 
We also showed that spin-coupling strengths (structural) fluctuations are less detrimental than on-site disorder since the former assures particle-hole symmetry and induces
the appearance of an eigenstate at the very center of the band showing ubiquitous localization properties. 

One of the advantages of such a class of weakly-coupled spin models offers is that
the communicating parties only have access to the spectrum of the channel locally through the
sites they are coupled to. Therefore, perfect spatial mirror symmetry
can be put aside as long as a proper set of delocalized states are available
in the spectrum, thus leading to an effective resonant coupling between both outer spins no matter how distant they are.




\acknowledgments
This work was partially supported by CNPq (Grant No. 152722/2016-5),
CAPES, FINEP, and FAPEAL (Brazilian agencies). 


\end{document}